\documentclass [12pt]{article}
\usepackage {graphicx}

\begin{document}
\begin{center}
\textbf{Vogel-Fulcher law of glass viscosity: A new approach}
\end{center}

\begin{center}
N. Kumar
\end{center}

\begin{center}
Raman Research Institute, Bangalore 560 080, India
\end{center}

Starting with an expression, due originally to Einstein, for the shear 
viscosity \textit{$\eta $}(\textit{$\delta \phi $}) of a liquid having a small fraction \textit{$\delta \phi $ }by volume of solid 
particulate matter suspended in it at random, we derive an effective-medium 
viscosity \textit{$\eta $}(\textit{$\phi $}) for arbitrary \textit{$\phi $} which is precisely of the Vogel-Fulcher form. An 
essential point of the derivation is the incorporation of the 
excluded-volume effect at each turn of the iteration \textit{$\phi $}$_{n + 1 = }$\textit{$\phi $}$_{n}$\textit{+$\delta \phi $}. 
The model is frankly mechanical, but applicable directly to soft matter like 
a dense suspension of microspheres in a liquid as function of the number 
density. Extension to a glass forming supercooled liquid is plausible 
inasmuch as the latter may be modelled statistically as a mixture of rigid, 
solid-like regions (\textit{$\phi $}) and floppy, liquid-like regions (1-\textit{$\phi $}), for \textit{$\phi $} increasing 
monotonically with supercooling.

PACS number(s): 64.70.Pf, 45.70.-n, 66.20.+d, 83.80.Hj

Extreme slow dynamics defines approach to the glassy state. At the 
macroscopic scale, it manifests as a rise of shear viscosity, typically by 
15 orders of magnitude, as that state is reached through supercooling of the 
glass forming liquid. The Vogel-Fulcher (VF) law describes that growth of 
viscosity [1].$^{ }$This work derives the VF law [2].

A striking feature of the VF law is the essential singularity, rather than a 
power-law divergence, of the shear viscosity at a temperature $T_{0}$. The 
relaxation times, however, exceed the experimental time scale at what is 
identified as the glass transition temperature $T_{g}>T_{0}$, making thus the 
glass transition a kinetic crossover. This inverse exponential VF law is well 
known to hold for the fragile structural-glass forming liquids [1]. But, 
very significantly it is also obeyed by a broad class of soft-matter systems 
that exhibit the extreme slow dynamics [1].$_{ }$ This includes the purely 
mechanical systems, e.g., of weakly perturbed granular aggregates, where the 
degree of compaction and the perturbation strength, rather than mass density 
and temperature, are the relevant variables and the control parameter, and 
the underlying physics is that of jamming, or blocking, by rigid granular 
contacts [3-6]. And, similarly for the case of a dense suspension of 
microspheres [1]. Motivated by its ubiquity and universality, we have 
attempted a derivation of the VF law for a fluid-mechanical model of a 
liquid containing a volume fraction \textit{$\phi $} of solid particulate matter suspended 
in it at random. It is an effective medium theory (EMT) along the line of 
Bruggemann's asymmetric EMT [7], but it goes beyond the mean field by 
incorporating the solid-solid exclusion explicitly in real space, which 
indeed is the essential point of our derivation. This, frankly mechanical 
model can, however, be re-interpreted as a model for the glass forming 
supercooled liquid inasmuch as the latter may be re-approximated as a 
statistical mixture of short-ranged rigidity (solid-like fraction \textit{$\phi $}) and the 
floppy liquid-like fraction (1-\textit{$\phi $}). In our view the present work complements 
the other derivations of the VF law which are based on the idea of marginal 
scaling [6] and some simple exclusion models [8-11].

We start with the expression, due originally to Einstein [12],$^{ }$for the 
shear viscosity \textit{$\eta $}(\textit{$\delta \phi $}) of a liquid containing a small volume fraction \textit{$\delta \phi $} of solid 
particulate matter suspended in it at random:

\begin{equation}
\label{eq1}
\eta (\delta \phi )\,\, = \,\,\eta (0)\,(1\, + \,\alpha \delta \phi )\,
\end{equation}

\noindent
where \textit{$\alpha $}, of order unity, is a fluid-dynamic dimensionless parameter 
specifying the particle shape and the flow boundary condition, and \textit{$\delta \phi $ = }(\textit{4$\pi $/3})$a^{3}$\textit{$\delta $n} 
assuming spherical particles of radius `$a'$ with \textit{$\delta $n} the number density. The 
physical basis of Eq. (\ref{eq1}) is that in the steady state \textit{the rigid parts of the liquid move practically as complete wholes, and hence the effect of their existence is to diminish the thickness of the layer, through which momentum has to be transported by the mobile molecules, and thus to increase the viscosity }[13]$.$ We can iterate 
Eq. (\ref{eq1}) to a higher volume fraction \textit{$\phi $,} in the spirit of an EMT, by the 
recursion relation

\begin{equation}
\label{eq2}
\eta (\phi + \delta \phi )\, = \,\eta (\phi )\,(1 + \alpha 
\,\textstyle{{\delta \phi } \over {1 - \phi }}),
\end{equation}

\noindent
where the factor 1/(1-\textit{$\phi $}) in the denominator on the right-hand side ensures 
that the elemental increment \textit{$\delta \phi $} is reckoned relative to the liquid-like volume 
fraction (1-\textit{$\phi $}) remaining at the current stage of iteration. Now, proceeding 
to the limit \textit{$\delta \phi  \to $}0, we obtain the differential equation

\begin{equation}
\label{eq3}
\frac{d\eta }{\eta }\,\, = \,\,\left( {\frac{\alpha }{1 - \phi }} 
\right)\,\,d\phi \,\,\,\,\,\,\,\,\,\,\,\,\,
\end{equation}

\noindent
with the solution

\begin{equation}
\label{eq4}
\eta (\phi )=  \eta (0)_ (1 - \phi )^{ - \alpha }
\end{equation}

\noindent
that gives a power-law divergence for the effective shear viscosity \textit{$\eta $}(\textit{$\phi $}). Here 
\textit{$\eta $}(0) is the `bare' viscosity of the pure liquid with \textit{$\phi $} = 0. Such a power-law 
temperature dependence is well known to follow from the viscosity feedback 
mechanism giving the Batchinski-Hildebrand law [14] (with \textit{$\alpha $}=1), or from the 
Mode Coupling Theory (MCT) [15] giving the critical behaviour (with \textit{$\alpha $}$ \cong 
$2). Both these exponent values lie in the range for the parameter \textit{$\alpha $} as 
described below.

Equation (\ref{eq4}) giving this critical behaviour is, however, \textbf{in error} 
in that it mathematically ignores the physically important excluded-volume 
effect. The point is that the liquid fraction (1-\textit{$\phi $}) in the denominator in Eq. 
(\ref{eq3}) must be replaced by the liquid fraction (1-\textit{$\phi $}) \textit{as weighted by the probability that the incremental solid fraction $\delta \phi $, added at random, }\textbf{\textit{lands}}\textit{ in it} 
[16]$. $This, therefore, effectively replaces (1-\textit{$\phi $}) by (1-\textit{$\phi $})$^{2. }$ Equation (\ref{eq4}) 
then gets modified accordingly to

\begin{equation}
\label{eq5}
\frac{d\eta }{\eta }\,\, = \,\,\frac{\alpha }{(1\, - \phi )^2}\,\,\,d\phi 
\end{equation}

\noindent
giving

\begin{equation}
\label{eq6}
\eta (\phi )\,\, = \,\,\left( {\eta (0)\,\,e^{ - \alpha \,}} 
\right)\,\,e^{\alpha / (1 - \phi )}.
\end{equation}

The expression in Eq. (\ref{eq6}) is already of the VF form as an inverse 
exponential function of \textit{$\phi $} diverging essentially at \textit{$\phi $ }=1. This, however, needs a 
refinement as dictated by the physics of the problem, namely, that the solid 
volume fraction \textit{$\phi $} need approach only the rigidity percolation threshold 
\textit{$\phi $}$_{0}$(<1) in order to attain the three-dimensional rigidity. Therefore, 
(1-\textit{$\phi $}) above must be displaced to (\textit{$\phi $}$_{0}$\textit{ - $\phi $}). Thus, we finally have

\begin{equation}
\label{eq7}
\eta (\phi ) = \left( {\eta_0 e^{ - \alpha /\phi_0}} 
\right)e^{\alpha / (\phi_0 - \phi )}.
\end{equation}

\noindent
which tends to \textit{$\eta $(0)} for \textit{$\phi  \to $}0 (pure liquid), and diverges as \textit{$\phi  \to \phi $}$_{0}$ from below (the 
glassy state).

Equation (\ref{eq7}) is our main result. For the simplest case of spherical, 
non-spinning particles, we have [12,13] \textit{$\alpha $} = 2.5, while for particles free to 
spin, \textit{$\alpha $}=1. Also, we can estimate the rigidity percolation threshold [17] 
$(\phi_0^{3D})$in three 
dimensions from its 2D value $\phi_0^{2D}\simeq 0.80$ by use of the simple relation $\phi_0^{3D} = 4/3\pi^{1/2}(\phi_0^{2D})^3/2$.  We get $\phi^{3D} \simeq$ 0.54. In Fig. (\ref{eq1}), we have plotted \textit{$\eta $}(\textit{$\phi $}) against \textit{$\phi $} for 
the values of the parameters, \textit{$\alpha $} = 2.5 and $\phi_0^{3D}$= 0.54. This is essentially a universal curve.

While Eq. (\ref{eq7}) is expected to be directly applicable to, e.g., a suspension 
of microspheres in a viscous liquid, its extension to the glass forming 
supercooled liquids is plausible as indicated earlier. Then \textit{$\phi $} must be 
regarded as a function of temperature, increasing monotonically as the 
temperature decreases. This will turn Eq. (\ref{eq6}) explicitly into the VF form, 
or its variant, the Vogel-Tammann-Fulcher (VTF) law, \textit{$\eta $}($T)$ = \textit{$\eta $}$_{0}$ exp(\textit{DT}$_{0}/(T-T_{0}))$ 
as \textit{T$ \to $ T}$_{0 }$ from above$.$

We would now like to conclude with the following remarks. The above 
fluid-mechanical model physically implies that our derivation may apply more 
readily to the fragile rather than to the strong (network forming) liquids. 
As noted above, the numerical value of \textit{$\alpha $} occurring in Eq. (\ref{eq7}) depends on the 
particle shape [12] (taken to be spherical here), and on whether the 
particles are free to spin (\textit{$\alpha $} = 1) or not (\textit{$\alpha $ }= 2.5) in the presence of a shear 
rate. This can make the parameter \textit{$\alpha $ }temperature dependent, with the higher 
value \textit{$\alpha $} = 2.5 appropriate to the lower temperatures. With the solid-like 
volume fraction \textit{$\phi $} now become a function of temperature, and, therefore, a 
thermodynamic parameter, the equation (\ref{eq7}) shows how the shear viscosity (a 
transport property) is actually controlled by the thermodynamics: The 
thermodynamically controlled liquid-like fraction (\textit{$\phi $}$_{0}$ -\textit{$\phi $}) acts as an 
\textit{idler} taking up the shear rate. This is the simplest realization of a viscosity 
amplification that underlies the macroscopic slow dynamics described by the 
Vogel-Fulcher law, where the idling liquid-like fraction essentially retains 
its `bare' low value \textit{$\eta $}(0).

\subsubsection{Acknowledgements}

This work is dedicated to the memory of late Prof. Sivaraj Ramaseshan who 
brought Prof. Raman's paper on the viscosity of liquids to my attention and 
encouraged me to pursue these ideas, even as his own health was failing.

\textbf{------------------}

E-mail: \underline {nkumar@rri.res.in}).

\newpage 
\begin{figure}[htbp]
\centerline{\includegraphics[width=7.50in,height=5.53in]{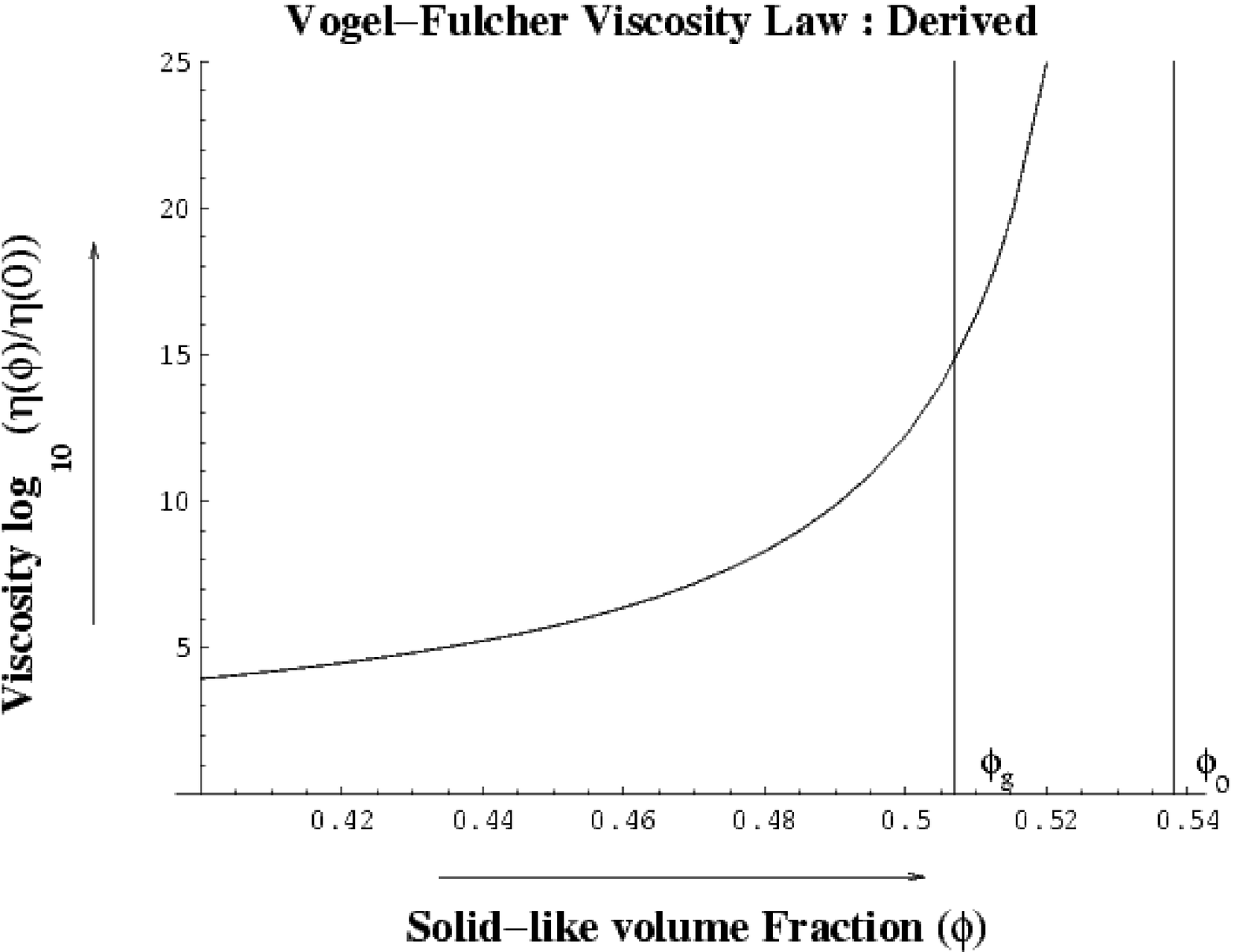}}
\label{fig1}
\end{figure}

FIG. 1. Plot of normalized shear viscosity \textit{$\eta $}(\textit{$\phi $})/\textit{$\eta $(0)} against the 
solid-like volume fraction \textit{$\phi $} from Eq. (\ref{eq7}) derived in the text, for \textit{$\alpha $} = 2.5. 
Here \textit{$\phi $}$_{0}$ is the rigidity percolation threshold, and \textit{$\phi $}$_{g}$ marks the 
point (\textit{$\phi $}$_{g})$/\textit{$\eta $(0)}$_{ }$= 10$^{15}$. The regime 0\textit{<$\phi $ <$\phi $}$_{g}$ is nominally the 
supercooled liquid; \textit{$\phi $}$_{g}$\textit{<$\phi $ <$\phi $(0)} the glassy liquid; and \textit{$\phi $>$\phi $}$_{0}$ the rigid glassy 
solid.

\end{document}